\documentclass[]{piparticle-final}
\usepackage{graphicx}
\usepackage{amsmath}

\usepackage{cite} 
\usepackage{epstopdf} 

\begin{document}

\volume{6}               
\articlenumber{060009}   
\journalyear{2014}       
\editor{L. A. Pugnaloni}   
\reviewers{L. Staron, CNRS, Universit\'e Pierre et Marie Curie, \\ \mbox{}\hspace{3.5cm} Institut Le Rond d'Alembert, Paris, France.}  
\received{{11} September 2014}     
\accepted{10 October 2014}   
\runningauthor{T. J. Wilson \itshape{et al.}}  
\doi{060009}         

\title{Granular discharge rate for submerged hoppers}

\author{T. J. Wilson,\cite{inst1,inst2}
	C. R. Pfeifer,\cite{inst1,inst3}
	N. Meysingier,\cite{inst1,inst4}
	D. J. Durian\cite{inst1}\thanks{Email: djdurian@physics.upenn.edu}}

\pipabstract{
The discharge of spherical grains from a hole in the bottom of a right circular cylinder is measured with the entire system underwater.  We find that the discharge rate depends on filling height, in contrast to the well-known case of dry non-cohesive grains.  It is further surprising that the rate increases up to about twenty five percent, as the hopper empties and the granular pressure head decreases.  For deep filling, where the discharge rate is constant, we measure the behavior as a function of both grain and hole diameters.  The discharge rate scale is set by the product of hole area and the terminal falling speed of isolated grains.  But there is a small-hole cutoff of about two and half grain diameters, which is larger than the analogous cutoff in the Beverloo equation for dry grains.
}

\maketitle

\blfootnote{
\begin{theaffiliation}{99}
	\institution{inst1} Department of Physics \& Astronomy, University of Pennsylvania, Philadelphia, PA 19104-6396, USA.
	\institution{inst2} Department of Physics, Illinois Wesleyan University, Bloomington, IL 61702-2900, USA.
	\institution{inst3} Department of Physics \& Astronomy, Carleton College, Northfield, MN 55057, USA.
	\institution{inst4} Strath Haven High School, Wallingford, PA 19086, USA.
\end{theaffiliation}
}


\section{Introduction}

The flow of granular materials is of widespread practical \cite{Nedderman, Muzzio02} and fundamental \cite{JNBrev96, DuranBook} interest.  An important example is the gravity-driven flow of grains from an hourglass or flat-bottomed hopper.  For ordinary fluids, the discharge rate is proportional to the pressure head, which is set by filling height, and decreases continuously to zero with vanishing hole size.  The discharge of grains is strikingly different.  In particular, for dry non-cohesive grains, the mass per unit time discharged from a hole in the bottom of a hopper is accurately described by the empirical Beverloo equation \cite{Beverloo}:
\begin{equation}
W=C\rho_b g^{1/2}(D-kd)^{5/2}.
\label{beverlooequation}
\end{equation}
Here $\rho_b$ is the mass density of the bulk granular medium, $g=980$~cm/s$^2$, $D$ is the hole diameter, $d$ is the grain diameter, and $C$ and $k$ are dimensionless fitting parameters \cite{Beverloo}.  The filling height plays no role.  The container diameter, and hence the grain-grain contact pressure at the bottom of the container, as given by the Janssen argument \cite{VanelEPJB1999}, also plays no role.  Intuitively, ``transient arches" intermittently form and break over the hole, shielding the exiting grains from any sort of pressure head.  Then the rough scale for discharge is the free-fall speed $\sqrt{g D}$ times hole area, i.e.,\ $W\sim \rho_b g^{1/2} D^{5/2}$.  Beverloo et al.\ plotted their data as $W^{2/5}$ versus $D$ and found a straight line, but such that $W$ vanishes at a nonzero small-hole cutoff diameter of $kd$ \cite{NeddermanSavage}. This has been rationalized by an ``empty annulus'' around the perimeter of the hole, through which grain centers cannot pass.  Nevertheless, 
fundamental 
justification of Eq.~(\ref{beverlooequation}) remains a topic of on-going interest \cite{HiltonPRE11, JandaPRL12}.  

The Beverloo Eq.~(\ref{beverlooequation}) is supported by a large number of experiments, as reviewed by Nedderman and Savage et al.~\cite{NeddermanSavage}.  But discrepancies of up to forty percent have been reported when the hole size is increased more widely than usual~\cite{MazaGM07}.  Recently, we have found excellent agreement with the Beverloo equation for up to three \cite{Hannah_GM10} and four \cite{Charles_PRE13} decades in discharge rate for spherical grains.  Typical ranges for the numerical coefficients are $0.5<C<0.7$ and $1.2<k<3$; near the low end for spherical grains.  Discharge rates can be increased, paradoxically, by placing an obstacle over the hole \cite{ZuriguelPRL11, LozanoPRE12, AlonsoPRE12}.  And for small enough holes, stable arches can form and cause a clog \cite{ToPRL01}.  Related behavior has been reported for the upward discharge of bubbles in an underwater silo \cite{BerthoPRE06}, particles on a conveyor belt \cite{AguirrePRL2010}, and for 
disks floating on a fluid that flows 
through an orifice \cite{GuariguataPRE2012, LafondPRE2013}.  An important challenge is to relate all such phenomena to the velocity and density fields, which can be measured in quasi-2D \cite{KudrolliPRL04, Kudrolli05, BehringerEPJE10, JandaPRL12, WeeksSM12, DenninPRE13} and index-matched \cite{KudrolliPRL07} systems.

The physical intuition behind the transient arch and empty annulus concepts is appealing, and helps explain the contrast between granular and Newtonian fluids.  But it has proven difficult to translate into a first-principles theory of hopper discharge and clogging.  This could be because the jamming and unjamming of grains in the converging flow near the boundary is not only collective, but is also even more difficult to model than jamming in uniform systems \cite{LiuNagel_ARCMP2010}.  As an experimental approach to alter transient arching and grain free-fall, we previously explored the effects of tilting the hopper and, hence, the plane of the hole away from horizontal \cite{Hannah_GM10, Charles_PRE13}.  Now in this paper, we perturb transient arching in a totally different way by submerging the entire system underwater.  Here the grain dynamics become overdamped rather than inertial, and the characteristic grain exit speed becomes set by the fluid and the grain size, rather than 
by free-fall and the hole 
size.  We find not only a change in the scale of Eq.~(\ref{beverlooequation}), but also in the small-hole cutoff.  This is our main focus.  In addition, we find an unexpected dependence of discharge rate on filling height, opposite in sign to that for ordinary liquids, which is to be the subject for further experiments \cite{Juha}.

\section{Materials and methods}\label{MM}

The granular media consist of monodisperse spherical beads, primarily of glass (Potter Industries), but also of lead (McMaster-Carr).  The average and standard deviation of the grain diameters $d$, the grain material density $\rho_g$, and the resulting terminal falling speed in water $v_t$, are collected in Table~\ref{grains}.  The latter are computed using an accurate empirical formula \cite{DragCoefficient} for the dimensionless drag coefficient $c_d$ versus Reynolds number ${\rm Re}=\rho v_t d/\eta$, where $\rho$ and $\eta$ are respectively the density and viscosity of water.  This is done by equating gravity and drag forces, $\Delta m g = c_d \rho {v_t}^2A/2$ where  $\Delta m = (4/3)(\rho_g-\rho)\pi(d/2)^3$ and $A=\pi(d/2)^2$, and solving numerically for $v_t$.  The packings are further characterized by the draining angle of repose, $\theta_r$, as measured in air and under water.  The results in Table~\ref{grains} show that $\theta_r$ is larger for glass beads when dry, but larger for lead beads when wet.
  Also, $\theta_r$ is noticeably larger for the $d=0.11$~mm glass beads.  Finally, note that the polydispersity is about 20\% for the $d=0.11$~mm glass beads and the lead beads, but is $1-6$\% for the other three glass bead samples.

\begin{table*}
\begin{center}
\begin{tabular}{lccccccc}
\hline
\hline
Grains   & $d$ (mm) &  $\rho_g$ (g/mL) & $v_t$ (cm/s) & Re & $\phi$ & $\theta_r$ dry (deg) & $\theta_r$ wet (deg) \\
\hline
glass &	$0.11\pm0.02$		& $2.499\pm0.007$	& 0.94	& 	1		& $0.57\pm0.01$	&	$29.8\pm0.2$ &	$26.5\pm0.3$ \\
glass &	$0.31\pm0.02$		& $2.504\pm0.003$	& 4.0		& 	12		& $0.59\pm0.01$	&	$24.8\pm0.3$ &	$22.7\pm0.3$ \\
glass &	$0.96\pm0.05$ 	& $2.519\pm0.007$	& 15		& 	150		& $0.62\pm0.01$	&	$24.6\pm0.3$ &	$21.5\pm0.3$ \\
glass &	$3.00\pm0.04$		& $2.590\pm0.010$	& 36		& 	1100 	& $0.57\pm0.01$	&	$24.2\pm0.3$ &	$19.8\pm0.8$ \\
lead  &	$1.0\pm0.2$ 		& $10.9\pm0.2$	& 49		& 	520		& $0.62\pm0.01$	&	$17\pm1$		&	$22\pm3$ \\
\hline
\hline
\end{tabular}
\end{center}
\caption{Granular materials properties: $d$ is the average grain diameter, $\rho_g$ is the density of individual grains, $v_t$ is the terminal falling speed of individual grains in water, Re is the corresponding Reynolds number, $\phi$ is the volume fraction of grains in the packing, $\theta_r$ is the draining angle of response as measured for both dry and submerged packings.  Measurement uncertainties are also given, except for the grain diameter where the standard deviation of the size distribution is reported.}
\label{grains}
\end{table*}

Two different hoppers types are used.  The first is constructed from 93~mm diameter transparent cylindrical plastic (polyethylene terephthalate) jars with screw-on plastic lids of equal diameter.  For these, an outlet hole of desired diameter $D$ is drilled into a plastic lid.  Nearly two dozen different hole diameters are used, each drilled into a different lid.  The hole diameters are measured by calipers with an uncertainty of $\pm0.02$~cm.  An inlet for water is opened on the other end of the jar in order to prevent a back-flow of water up into the hopper to replace the lost volume of discharged grains.  The other hopper type consists of plastic graduated cylinders, diameter 50~mm or 38~mm, with a single hole drilled into the bottom.

All discharge measurements are conducted with the hopper fixed to a sturdy aluminum stand, all completely underwater in a large aquarium.  Prior to use, the glass beads are submerged and repeatedly poured back and forth between two containers in the same aquarium in order to allow all air to escape.  The grains are then poured slowly into the hopper, without exposure to air, and allowed to settle with the outlet hole blocked.  As such, the packings have a solids volume fraction $\phi$ near random-close packing (Table~\ref{grains}).  For large holes, flow commences immediately after the hole is unblocked.  For small holes, gentle tapping is required to start the discharge.  In either case, the discharge rate is measured only after the flow has proceeded long enough that a conical depression fully develops at the top of the packing.  The height $H$ of the packing is then measured as the level of the grains at the container wall.  The height of the packing over the 
hole is then estimated as $h=H-(D/2)\tan\theta_r$.

The mass per unit time discharge rate, $W$,  is found by starting/stopping a timer while inserting/removing a cup from the discharge stream.  The collection cup is then removed from the aquarium and weighed, carefully topped off with water.  The mass of the same cup, filled with water but no grains, is also determined.  The difference between the two mass measurements is $\Delta M=V_g(\rho_g-\rho)$ where $V_g$ is the volume of grains.  This gives the mass of the collected grains as $M_g=V_g \rho_g = \Delta M \rho_g/(\rho_g-\rho)$.  The discharge rate, $W$, is then computed as $M_g$ divided by the duration of the collection time.  The collection times range from a few seconds to several hours, and $M_g$ ranges from about 10~g to 250~g for the glass beads.  Collection of grains is  begun only after a steady-state depression is formed at the top surface of packing; thus the possibility of initial transients, which could depend on initial packing fraction \cite{RondonPF2011}, is not investigated.

\section{Versus Packing Height}

\begin{figure}
\includegraphics[width=3.00in]{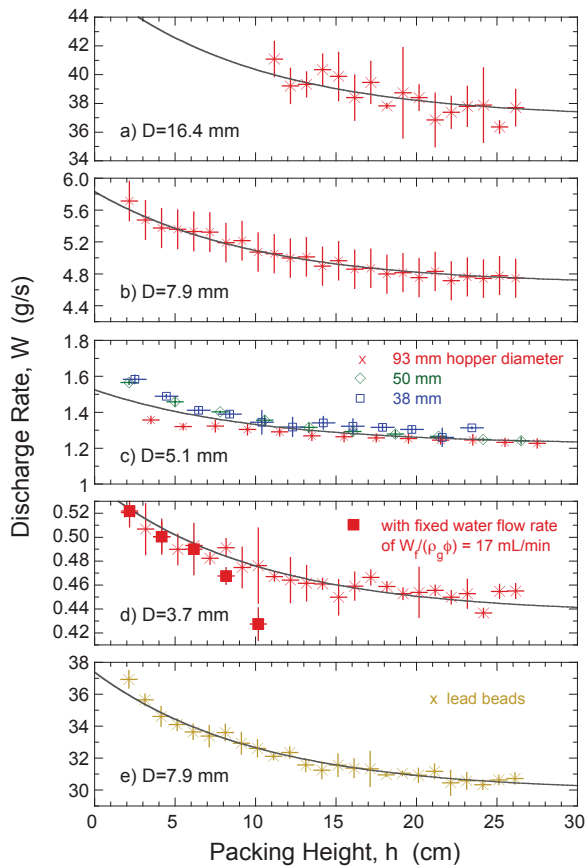}
\caption{Discharge rate vs packing height for (a-d) $d=1$~mm diameter glass beads and (e) $d=1$~mm diameter lead beads, and different hole diameters $D$ as labeled.  Note that time proceeds right to left.  The hopper diameter is 93~mm, except as noted in (c).  In (d), data are also included where the container is sealed and a fixed flow of water is pumped into the top at the same volumetric rate at which grains exit the hole at bottom.  The horizontal error bars indicate the range of packing heights over which the discharge rate is measured.  The vertical error bars indicate the standard deviation of 3-5 repetitions.  The solid curves are fits to $W(h)=W_o\{1+0.25\exp[-h/(10~{\rm cm})]\}$, where $W_o$ is the only adjustable parameter, to show that the height-dependence is similar for different discharge geometries.}
\label{Wvsh}
\end{figure}

A prudent preliminary task is to measure whether or not the discharge rate $W$ for submerged grains is independent of the packing height $h$, as in the case of dry grains.  So in Fig.~\ref{Wvsh} we show $W$ versus $h$ data for $d=1$~mm beads, of glass and lead, for holes of various diameter.  The rates all appear to approach a constant for very large packing heights.  Therefore, the driving pressure on the grains at the outlet must be shielded from the weight of the packing due to transient arch formation as for dry grains.  For smaller packing heights, however, the discharge rate data in Fig.~\ref{Wvsh} are not constant.  Rather, the rates all {\it increase} as the hopper empties.  This is surprising, because the sign of the effect is opposite to that for ordinary fluids where the rate decreases with the diminishing pressure head as the container empties.  Similar behavior has been noticed for submerged hoppers \cite{Arshad}, and also for dry grains in a cylindrical tube with a conical orifice \
cite{LePennecPF98}.  Preliminary data, with a new apparatus, show that the discharge increase upon emptying is much smaller for dry grains, on the order of one percent \cite{Juha}.  We are unaware of any theoretical explanation.  In fact, both the $\mu(I)$ flow law and discrete contact dynamics simulations for two dimensional dry grains predict a decrease in discharge rate as the hopper empties \cite{StaronPF12, Staron2013}.  The former approach has been extended to suspensions \cite{BoyerPRL2011}, but not yet applied to hopper flows.

The variation of discharge rate with packing height is roughly exponential.  Fits to the form $W(h)=W_o\{1+0.25\exp[-h/(10~{\rm cm})]\}$ are shown by solid curves in Fig.~\ref{Wvsh}, where the asymptotic discharge rate $W_o$ is the only fitting parameter.  This function has no theoretical basis, as yet, but is a simple form that allows us to both estimate $W_o$ and to compare the displayed data sets.  The fits are quite good, and thus illustrate (i) that the size of the surge is about 25\% at most, and (ii) that the characteristic height is about 10~cm.  These two numbers represent the average of individual fitting results for all hole sizes for the 93~mm hoppers, when all three parameters are allowed to float.  For the 93, 50 and 38~mm diameter hoppers, shown in Fig.~\ref{Wvsh}(c), the individually-fitted decay lengths are $14.6\pm3$, $8.7\pm0.4$ and $4.9\pm0.8$~cm, respectively.  Thus, the decay length appears to be on the order of ten times the hopper diameter.  A factor of order one 
might have been 
expected 
if Janssen-type wall effects were primarily responsible.  Nonetheless, one hypothesis for the surge would be that the grain-grain contact pressure over the outlet decreases as the hopper empties, and a resulting subtle decrease in packing fraction allows for greater fluidity and hence greater discharge rate.  However, this seems ruled out by Ref.~\cite{PergePRE2012}, where data show a decrease of pressure but no change in discharge rate.

Whatever the cause of the packing height dependence of discharge rate, the fluid clearly plays a role.  So it is logical to consider interstitial flow between the grains.  Certainly, such flow is generated as the grains move apart and out of the hole.  A second hypothesis would be that some liquid is drawn down through the whole packing, even more easily and rapidly as the hopper empties ---which would cause the grain discharge rate to increase.  For large packing heights, such fluid flow would vanish and the grain discharge rate would become constant.  As a test, we sealed off the top of a 93~mm diameter hopper and connected it to a gear pump for enforcing a constant interstitial liquid flow rate.  The liquid pump rate was set to $W_o/(\rho_g \phi)$, so that grains and fluid can flow inside the hopper together, in unison, with the same macroscopic velocity field.  This would be the condition for very tall packings, but is now enforced for all 
packing heights.  
Grain discharge rate data are plotted in Fig.~\ref{Wvsh}(d), on top of prior data for an open-topped hopper with no liquid pumping.  Except for one perhaps-spurious point, the two data sets are indistinguishable in showing the same rate increase as the hopper empties.  Therefore, interstitial fluid flow down through the packing is not responsible for the surprising variation of discharge rate with packing height.  The fluid and the walls probably play a role, but the mechanism is not known.  Further experiments are now in progress to address this issue \cite{Juha}.

\section{Versus Hole Diameter}

As our main task, we now turn to the variation of discharge rate with the hole diameter $D$, in the limit of large packing heights where the rate has a constant value $W_o$.  For this, we collect data for 93~mm diameter hoppers with packing heights over the range $25~{\rm cm}<H<28~{\rm cm}$; this procedure agrees well with results from exponential fits illustrated in Fig.~\ref{Wvsh}.  The simplest scale for discharge rate is $W_o \propto \rho_g v_t D^2$, where $\rho_g$ is the density of the grain material, $v_t$ is the terminal falling speed for an individual grain in water.  Note that $v_t$ depends on fluid and grain properties, and scales with grain diameter as $d^2$ at low Re but as $d^{1/2}$ at high Re (see Section~\ref{MM}).  This contrasts with dry grains, where the discharge speed depends on grain size only for very small holes.   Imposing a Beverloo-like cutoff by the substitution $D\rightarrow (D-kd)$, we arrive at the expectation
\begin{eqnarray}
	W_o &=& C \rho_g v_t (D-kd)^2, \label{rate1} \\
	        &=& C (\rho_g v_t d^2)[(D/d)-k]^2.\label{rate2}
\end{eqnarray}
Similar Beverloo-type forms with cutoff have been found for the upward discharge of bubbles in a quasi-2D silo \cite{BerthoPRE06} and for particles on a conveyor belt \cite{AguirrePRL2010}.  Consideration of $v_t$ based on a single grain warrants caution, since there must be a more complex flow of the interstitial fluid in response to granular shear and dilation near the aperture.

\begin{figure}
\includegraphics[width=3.00in]{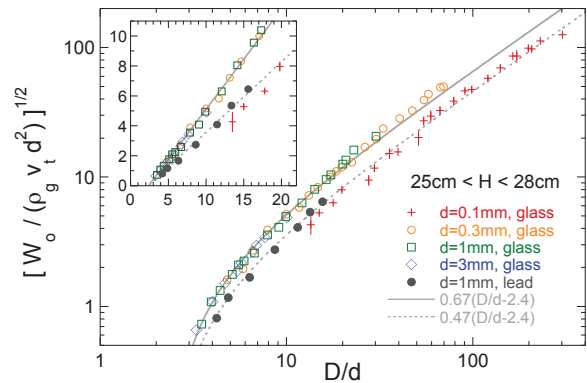}
\caption{Mass discharge rate $W_o$ vs hole diameter $D$, in the large packing-height limit, for several grain types as labelled.  The axes are made dimensionless by grain diameter $d$, density $\rho_g$ of individual grains, and terminal falling speed $v_t$ of individual grains in water.  The packing height range is $25~{\rm cm}<H<28~{\rm cm}$, and the hopper diameter is 93~mm.  Vertical error bars, given by the standard deviation of 3-5 repetitions, are smaller than symbol size except for the $d=0.1$~mm glass beads as shown.  The solid curves represent lines that vanish at $D/d=2.4$, as specified in the legend.  Fits to ${W_o}^{1/2} \propto (D/d-k)$ give the average cutoff as $k=2.4\pm0.1$ grain diameters.}
\label{WvsD}
\end{figure}

For comparison of data with Eq.~(\ref{rate2}), Fig.~\ref{WvsD} shows a dimensionless plot of discharge rate measurements as $[W_o/(\rho_g v_t d^2)]^{1/2}$ versus $D/d$.  Indeed, for glass beads of diameter 0.3, 1, and 3~mm, we find excellent collapse of the data onto a single line that vanishes at non-zero $D/d$.  This analysis is analogous to the famous Beverloo plot of $W^{2/5}$ versus $D/d$ for dry grains~\cite{NeddermanSavage}.  Here the main plot is double-logarithmic to illustrate a range of roughly two decades in dimensionless hole diameter and five decades in dimensionless discharge rate; the inset has linear axes with a smaller range so that the functional form versus hole diameter is evident.  Data for the lead and the 0.1~mm glass beads show similar linear behavior, but do not collapse onto the results for the larger glass beads.  Perhaps this is because their polydisersities are significantly larger, as seen in Table~\ref{grains}.

Two lines are shown in Fig.~\ref{WvsD} that correspond to Eq.~(\ref{rate2}) and match the two groupings of data.  The proportionality constants, $C$, are close to 2/3 and 1/2. It is reassuring that these values are of order one.  But it is more interesting that the lines both vanish at the same ratio of hole to grain diameter.  Based on individual linear fits to the separate data sets, this small-hole cutoff is $k=2.4\pm0.1$.  By contrast, for dry spherical grains, the Beverloo cutoff for Eq.~(\ref{beverlooequation}) is $k=1.5$; and for air bubbles in water it is $k=0.66$.  Therefore, the cutoff involves dynamics and cannot be explained by purely geometrical concepts like the ``empty annulus''.  Alternatively, our data could also be taken as further evidence for the point of view that discharge rate should be described by a functional form in which $k=1$ is enforced \cite{MazaGM07, JandaPRL12}.

\section{Conclusion}

In summary, we have observed a striking dependence of the discharge rate on packing height, such that the rate surges as the hopper empties.  The mechanism is not yet understood, but may involve wall effects and the flow of liquid in between grains as they come apart near the exit.  The latter is reminiscent of an increase of flux due to dilation of grains under a fixed obstacle placed above the hole \cite{ZuriguelPRL11, LozanoPRE12, AlonsoPRE12}.  But here the relevant length scale appears to be the hopper diameter, rather than the hole diameter, and the effect vanishes for very tall packing heights where the discharge rate is constant.  In this regime, for two spherical grain types and a wide range of grain diameters, we find excellent agreement between discharge rate data and a modified Beverloo-like equation over five decades in dimensionless discharge rate.  This empirical result could be of practical use, in the same vein as the Beverloo equation.  While the basic scale is set by dimensional 
analysis, 
we find the existence of a small-hole cutoff such that the discharge rate extrapolates to zero as the hole diameter is decreased to 2.4 times the grain diameter.  For dry grains, the cutoff is significantly smaller; this has been known for over fifty years but remains unexplained based on the underlying microscopic physics.  The contrast with our findings suggests a missing ingredient: grain dynamics, rather than geometry alone, play a crucial role.

\begin{acknowledgements}
We thank Ted Brzinski, Charles Thomas, and Adam Roth for helpful discussions and assistance.  This work was supported by the NSF through grants DMR-1305199 and MRSEC/DMR-112090 (REU program - TJW and CRP).
\end{acknowledgements}


\end{document}